

\newcount\instnumber  \instnumber=0
\def\instfoot#1{\global\advance\instnumber by1
  \xdef#1{$^{(\the\instnumber)}$}}
\catcode`@=11 
%
\def\space@ver#1{\let\@sf=\empty \ifmmode #1\else \ifhmode
   \edef\@sf{\spacefactor=\the\spacefactor}\unskip${}#1$\relax\fi\fi}
\newdimen\authrefminspace  \authrefminspace=2pc
\newcount\authrefcount     \authrefcount=96
\newif\ifauthreferenceopen       \newwrite\authreferencewrite
\newtoks\authrw@toks
\let\PRauthrefmark=\attach
\def\authrefmark#1{\relax\PRauthrefmark{#1}}
\newcount\lastauthrefsbegincount \lastauthrefsbegincount=0
\def\authrefsend{\authrefmark{\count255=\authrefcount
   \advance\count255 by-\lastauthrefsbegincount
   \ifcase\count255 \number\authrefcount
   \or \number\lastauthrefsbegincount,\number\authrefcount
   \else \number\lastauthrefsbegincount-\number\authrefcount \fi}}
\def\authrefch@ck{\chardef\authrw@write=\authreferencewrite
   \ifauthreferenceopen \else \authreferenceopentrue
   \immediate\openout\authreferencewrite=authrefe.texauxil \fi}
{\catcode`\^^M=\active 
  \gdef\obeyendofline{\catcode`\^^M\active \let^^M\ }}%
{\catcode`\^^M=\active 
  \gdef\ignoreendofline{\catcode`\^^M=5}}
{\obeyendofline\gdef\authrw@start#1{\def\t@st{#1} \ifx\t@st\blankend%
\endgroup \@sf \relax \else \ifx\t@st\bl@nkend \endgroup \@sf \relax%
\else \authrw@begin#1
\backtotext
\fi \fi } }
{\obeyendofline\gdef\authrw@begin#1
{\def\n@xt{#1}\authrw@toks={#1}\relax%
\authrw@next}}
\newif\iffirstauthrefline  \firstauthreflinetrue
\def\rwr@teswitch{\ifx\n@xt\blankend \let\n@xt=\authrw@begin %
 \else\iffirstauthrefline \global\firstauthreflinefalse%
\immediate\write\authrw@write{\noexpand\obeyendofline \the\authrw@toks}%
\let\n@xt=\authrw@begin%
      \else\ifx\n@xt\authrw@@d \def\n@xt{\immediate\write\authrw@write{%
        \noexpand\ignoreendofline}\endgroup \@sf}%
             \else \immediate\write\authrw@write{\the\authrw@toks}%
             \let\n@xt=\authrw@begin\fi\fi \fi}
\def\authrw@next{\rwr@teswitch\n@xt}
\def\authrw@@d{\backtotext} \let\authrw@end=\relax
\let\backtotext=\relax
\newdimen\authrefindent     \authrefindent=17pt
\def\authrefitem#1{\par \parskip=0pt
\hangafter=0 \hangindent=\authrefindent \Textindent{#1}}
\def\authrefnum#1{\space@ver{}\authrefch@ck \firstauthreflinetrue%
 \global\advance\authrefcount by 1 \xdef#1{\the\authrefcount}}

\def\authref#1{\authrefnum#1%
 \immediate\write\authreferencewrite{\noexpand\authrefitem{(\char#1)}}%
\begingroup\obeyendofline\authrw@start}
%
%
%
\catcode`@=12 
%

\PHYSREV

\def\etal{{\it et al.}}

\def\u4s{\Upsilon_{4S}}

\def\z0{$Z^0$}

\def\alp{\relax\ifmmode \alpha_s\else $\alpha_s$\fi$\;$}
\def\alpmz{\relax\ifmmode \alpha_s(M_Z)\else $\alpha_s(M_Z)$\fi$\;$}
\def\alpmzsq{\relax\ifmmode \alpha_s(M_Z^2)\else
    $\alpha_s(M_Z^2)$\fi$\;$}

\def\zbbar{Z^0 \rightarrow b \overline{b}}
\def\zccbar{Z^0 \rightarrow c \overline{c}}
\def\zuds{Z^0 \rightarrow uds}

\def\zhad{Z^0 \rightarrow hadrons}
\def\nb{\overline{n}_{b}}
\def\nc{\overline{n}_{c}}
\def\nhad{\overline{n}_{had}}
\def\nuds{\overline{n}_{uds}}
\def\db{\delta\overline{n}_b}
\def\mh{\overline{m}_{h}}
\def\mt{\overline{m}_{t}}
\def\nudsc{\overline{n}_{udsc}}

\def\nnl{\overline{n}_{nl}}
\def\ndk{\overline{n}_{dk}}

\overfullrule=0pt
 \hsize=6in
   \vsize=8.5in
\baselineskip = 16pt plus 1.5pt minus 1pt

\overfullrule=0pt
\linepenalty=100
\uchyph=200
\brokenpenalty=200
\hbadness=10000
\clubpenalty=10000
\widowpenalty=10000
\displaywidowpenalty=10000
disp
\pretolerance=10000
\tolerance=2000
\nobreak
\penalty 5000
\hyphenpenalty=5000
\exhyphenpenalty=5000
\sequentialequations


\abovedisplayskip 8pt
\abovedisplayshortskip 6pt
\belowdisplayskip 7pt


\frontpagetrue
\doublespace

\title{MEASUREMENT
OF THE CHARGED MULTIPLICITY OF $\zbbar$ EVENTS}

 %
 %
 %
 \instfoot\iADEL
 \instfoot\iBU
 \instfoot\iBRUN
 \instfoot\iCIT
 \instfoot\iCOL
 \instfoot\iFRA
 \instfoot\iBOL
 \instfoot\iFER
 \instfoot\iPAD
 \instfoot\iPERU
 \instfoot\iPISA
 \instfoot\iLBL
 \instfoot\iMIT
 \instfoot\iNAG
 \instfoot\iRUT
 \instfoot\iRAL
 \instfoot\iSLAC
 \instfoot\iTOH
 \instfoot\iUBC
 \instfoot\iUCSB
 \instfoot\iUCSC
 \instfoot\iCIN
 \instfoot\iCOLO
 \instfoot\iGEN
 \instfoot\iILL
 \instfoot\iMASS
 \instfoot\iOREG
 \instfoot\iUPERU
 \instfoot\iTENN
 \instfoot\iUVIC
 \instfoot\iWASH
 \instfoot\iWISC
 \instfoot\iVAND
 \instfoot\iYALE
 %
 %
 \author{\twelvepoint\singlespace
\hbox{K. Abe                 \unskip,\iTOH}
\hbox{I. Abt                 \unskip,\iILL}
\hbox{W.W. Ash               \unskip,\iSLAC}
\hbox{D. Aston               \unskip,\iSLAC}
\hbox{N. Bacchetta           \unskip,\iPAD}
\hbox{K.G. Baird             \unskip,\iRUT}
\hbox{C. Baltay              \unskip,\iYALE}
\hbox{H.R. Band              \unskip,\iWISC}
\hbox{M.B. Barakat           \unskip,\iYALE}
\hbox{G. Baranko             \unskip,\iCOLO
% ...
}
 }

\centerline{(The SLD Collaboration)}

\medskip
 %
 \address{\singlespace
 \iADEL
 Adelphi University,
 Garden City, New York 11530 \break
 \iBU
 Boston University,
 Boston, Massachusetts 02215 \break
 \iBRUN
 Brunel University,
 Uxbridge, Middlesex UB8 3PH, United Kingdom \break
 \iCIT
 California Institute of Technology,
 Pasadena, California 91125 \break
 \iCOL
 Columbia University,
 New York, New York 10027 \break
 \iFRA
 INFN  Lab. Nazionali di Frascati,
 I-00044 Frascati, Italy \break
 \iBOL
 INFN Sezione di Bologna,
 I-40126 Bologna, Italy \break
 \iFER
 INFN Sezione di Ferrara and Universit\`a di Ferrara,
 I-44100 Ferrara, Italy \break
 \iPAD
 INFN Sezione di Padova and Universit\`a di Padova,
 I-35100 Padova, Italy \break
 \iPERU
 INFN Sezione di Perugia and Universit\`a di Perugia,
 I-06100 Perugia, Italy \break
 \iPISA
 INFN Sezione di Pisa and Universit\`a di Pisa,
 I-56100 Pisa, Italy \break
 \iLBL
 Lawrence Berkeley Laboratory, University of California,
 Berkeley, California 94720 \break
 \iMIT
 Massachusetts Institute of Technology,
 Cambridge, Massachusetts 02139 \break
 \iNAG
 Nagoya University,
 Chikusa-ku, Nagoya 464 Japan  \break
 \iRUT
 Rutgers University,
 Piscataway, New Jersey 08855 \break
 \iRAL
 Rutherford Appleton Laboratory,
 Chilton, Didcot, Oxon OX11 0QX United Kingdom \break
 \iSLAC
 Stanford Linear Accelerator Center,
 Stanford, California 94309 \break
 \iTOH
 Tohoku University,
 Sendai 980 Japan \break
 \iUBC
 University of British Columbia,
 Vancouver, British Columbia V6T 2A6 Canada \break
 \iUCSB
 University of California at Santa Barbara,
 Santa Barbara, California 93106 \break
 \iUCSC
 University of California at Santa Cruz,
 Santa Cruz, California 95064 \break
 \iCIN
 University of Cincinnati,
 Cincinnati, Ohio 45221 \break
 \iCOLO
 University of Colorado,
 Boulder, Colorado 80309 \break
\iGEN
Universit\'a di Genova, 1-16146, Genova, Italy
 \iILL
 University of Illinois,
 Urbana, Illinois 61801 \break
 \iMASS
 University of Massachusetts,
 Amherst, Massachusetts 01003 \break
 \iOREG
 University of Oregon,
 Eugene, Oregon 97403 \break
\iUPERU
Universit\'a di Perugia,
 I-06100 Perugia, Italy \break
 \iTENN
 University of Tennessee,
 Knoxville, Tennessee 37996 \break
 \iUVIC
 University of Victoria,
 Victoria, British Columbia V8W 3P6 Canada \break
 \iWASH
 University of Washington,
 Seattle, Washington 98195 \break
 \iWISC
 University of Wisconsin,
 Madison, Wisconsin 53706 \break
 \iVAND
 Vanderbilt University,
 Nashville, Tennessee 37235 \break
 \iYALE
 Yale University,
 New Haven, Connecticut 06511 \break
 }
\vskip 0.3in
 %
\centerline{\bf Abstract}

\doublespace

\noindent
Using an impact parameter tag to select an enriched sample of $\zbbar$
events, we have measured the difference between the average
charged multiplicity of $\zbbar$ and $Z^0 \rightarrow hadrons$
to be $\nb - \nhad =
2.24\pm 0.30(\rm{stat.}) \pm 0.33(\rm{syst.})$ tracks per event.
From this, we have derived $\nb - \nuds = 3.31 \pm 0.41 \pm 0.79.$
Comparing this measurement with those at lower center-of-mass
energies, we find no evidence that $\nb - \nuds$ depends on energy.
This result is in agreement with a precise prediction
of perturbative QCD, and
supports the notion that QCD remains asymptotically free down to
the scale
$M_b^2$.

\vfill

\eject

\doublespace

Heavy quark systems are a
particularly good laboratory for detailed studies of the strong
interaction and tests of the theory of
Quantum Chromodynamics (QCD).
The
large quark mass $M_Q \gg \Lambda_{QCD},$ where $\Lambda_{QCD}$
is the QCD interaction scale, provides a natural
cutoff in the parton shower evolution,
which keeps the relevant space-time region compact enough
to avoid the non-perturbative domain of the strong
interaction. Recently it has been recognized
that, within the context of perturbative QCD,
this cutoff allows a stringent constraint to be placed on the
difference in light hadron production between
$e^+e^-$
annihilation into heavy and light quarks [1].
In particular, it is expected that to
$\displaystyle{O} ([\alpha_s(W^2)]^{1/2}(M_Q^2/W^2))$
($\simeq 0.1$ track at $W=M_Z$),
the difference between the
total mean charged multiplicity in light quark ($q=u,d,s$) events
and the
mean charged multiplicity of radiated `non-leading'
hadrons in heavy quark ($Q=b,c$) events, excluding the
decay products
of the `leading' long-lived heavy hadrons,
should be {\it independent} of center-of-mass
(cms) energy $W$. This is a striking prediction, in that the total
multiplicity is known to grow faster than logarithmically with W.
Furthermore, to
$\displaystyle{O} (\alpha_s(M_Q^2)\nuds(M_Q))$
($\simeq 1.2$ tracks for $Q=b$),
this multiplicity difference should be equal to
$\nuds(\sqrt{e}M_Q)$, the mean charged
multiplicity for $e^+e^-$ annihilation to light quarks at the
reduced cms energy $\sqrt{e}M_Q$, where $\ln{e} = 1$.
A test of this
hypothesis provides the opportunity to verify an accurate prediction of
perturbative QCD, and to probe the validity of perturbative
calculations
down to the scale $M_Q^2$.
In addition, this hypothesis is in direct contradiction with
the hypothesis of
flavor-independent fragmentation [2,3], which suggests
that the non-leading multiplicity associated with
heavy quark production at a given cms energy $W$ should be
equal to the total light quark ($u,d,s$) event multiplicity at the
reduced cms energy $(1- \langle x_Q \rangle)W$, where
$x_Q = 2 \cdot E_Q/W$ is the heavy hadron energy fraction after
fragmentation.

Recent tests of these hypotheses [1,4] made use of a
measurement of the mean charged multiplicity of $\zbbar$ events
from the statistically-limited data sample of the 1990 run of the
Mark II detector at the SLAC Linear Collider (SLC), and were
not able to demonstrate a clear preference for either.
Here, we present a more accurate
measurement based on the 1992 run of the SLC Large Detector
(SLD) experiment, during which a
total of
420 $nb^{-1}$ of
electron-positron
annihilation data were recorded at a mean cms energy of $91.55$ GeV.

The SLD is a multi-purpose particle detector and is described
elsewhere [5].
Charged particles are tracked and momentum analyzed
in the Central Drift Chamber (CDC), which
consists of 80 layers of axial or stereo sense wires in a
uniform axial magnetic field of 0.6T. In addition, a silicon
vertex detector (VXD) [6]
provides an accurate measure
of particle trajectories close to the beam axis.
With the exception of the hadronic event trigger, this analysis relied
exclusively upon the information from these two tracking systems.

While the multiplicity measurement relied primarily
on information from the CDC, the more accurate impact
parameter measurement provided by the addition of the VXD
information to the CDC tracks was used to select a sample
enriched in $\zbbar$ events. All impact parameters used in this
analysis were for tracks projected into the plane perpendicular
to the beam axis, and were measured with respect to an
average primary
vertex (PV) derived from fits to events
close in time to the event under study.
The impact parameter
$d$ was derived
by applying a sign to the distance of closest approach such
that $d$ is positive when the vector from the PV to the point
at which the track intersects the thrust axis [7]
makes an acute angle
with respect to the track direction.
Including the
uncertainty on the average PV, the measured
impact parameter uncertainty $\sigma_d$
for the overall tracking system approaches 15~$\mu m$ for
high momentum tracks, and is 80~$\mu m$ at $p_{\perp}\sqrt{\sin\theta}
= 1$ GeV/c, where $p_{\perp}$ is the momentum transverse to the
beam axis, and $\theta$ the angle relative to the beam axis.

Events were classified as hadronic decays of the $Z^0$
provided that they contained at least 7 tracks which intersected
a cylinder of radius
$r_0 = 5$ cm and half-length $z_0 = 10$ cm surrounding the average PV,
a visible charged energy of least 20 GeV,
and a thrust axis satisfying
$|\cos\theta_{thrust}| < 0.7$. The resulting sample contained
5449 events.
Backgrounds in this sample were estimated to be $\sim0.1\%$.

For the purpose of multiplicity counting, a loose set
of requirements was placed on reconstructed tracks,
while stricter requirements were placed on tracks used
to measure impact parameters.
`Multiplicity quality' tracks were
required to: i) have
$p_{\perp} \geq 0.12$~GeV/c; ii) have
$|\cos\theta|~\leq~0.8$; and iii) intersect a cylinder of
$(r_0,z_0)~=~(1.5,5.0)$~cm.
`Impact parameter quality' tracks were required to: i) have
$|\cos \theta|~\leq~0.8$; ii) intersect a cylinder of
$(r_0,z_0)~=~(0.3,1.5)$~cm; iii) have at least one VXD hit;
iv) have $\sigma_d < 250 \mu m$; and
v) have $\chi ^2/d.o.f.$ for the CDC-only and combined CDC/VXD fits
of less than 5.0 and 10.0, respectively.

A $\zbbar$ enriched sample was selected by dividing each event into
two hemispheres separated by the plane perpendicular
to the thrust axis, and requiring two
or more impact parameter quality tracks in one
hemisphere
with normalized impact parameter $d/\sigma_{d}>3.0$ [8].
Restricting the tag to tracks from a single hemisphere allowed
potential tagging bias to be reduced
by measuring the multiplicity in the hemisphere
opposite to the tag. Monte Carlo (MC) studies indicate that this
tag is 50\% efficient at identifying hemispheres containing
$B$ hadrons in selected hadronic events,
while providing an enriched sample of 72\% purity.
The tag selected 1829 hemispheres.

In determining the total charged $\zbbar$ multiplicity $\nb$,
we minimized
systematic errors
by measuring $\db \equiv \nb - \nhad$, and then adding
back in the total hadronic charged multiplicity $\nhad$, which
has been accurately determined by other experiments [9].
In terms of the
{\it uncorrected} mean reconstructed multiplicities $\mh$ ($\mt$)
of the total hadronic (hemisphere opposite tag) samples [4],
$$\db = (1-R_b)(\ndk + \nnl - \nudsc),$$
where $\nnl$ and $\nudsc$ satisfy
$$ \mh = C_{h,udsc}(1-P_h)\nudsc + C_{h,dk}P_h \ndk +
         C_{h,nl} P_h \nnl,$$
$$ 2\mt = C_{t,udsc}(1-P_t)\nudsc + C_{t,dk}P_t \ndk +
         C_{t,nl} P_t \nnl,$$
and where $P_h$ and $P_t$ are the fraction of $\zbbar$ events
 in the hadronic and
tagged samples, determined by MC studies to be 0.223 and 0.724,
respectively.
We have used the Standard Model value
$R_b = \Gamma(\zbbar)/\Gamma(Z^0 \rightarrow hadrons) = 0.217$ [10].
We have separated the $\zbbar$ multiplicity into two components,
one associated with the decay of the $B$ hadrons ($dk$), and one
associated with the remaining
non-leading system ($nl$), in order to take
advantage of measurements from the $\u4s$ which constrain both the
multiplicity and spectrum of $B$ hadron decay products [11,12].
Here $\ndk = 10.88 \pm 0.22$ is
twice the $B$ hadron decay multiplicity
from the $\u4s$ [11], with
an additional uncertainty of $\pm 0.10$ tracks
included to account for the uncertainty in the
production fractions and decay multiplicities of the $B_s$ and
$B$ baryons.
The constants $C_{i,j}$ account for the
effects of detector acceptance and inefficiencies, and biases
introduced by the event and tagged sample selection criteria.
The $C_{i,j}$ were
evaluated, using a MC simulation of the detector,
as the ratio of the
number of multiplicity quality
tracks to generated charged multiplicity
tracks for the six sub-samples.
We have included in the generated multiplicity any charged
track which is prompt, or is the decay product of a particle
with mean lifetime less than $3 \times 10^{-10}$~s.

Because of the exclusion of tracks with very low momentum or large
$|\cos \theta|$, the constants $C_{i,j}$ are somewhat dependent
on the model used to generate MC events; we have used
JETSET 6.3 [13] with parameter values
tuned to hadronic $e^+e^-$ annihilation data [14].
The resulting values
for the $C_{i,j}$ were 0.855, 0.905, and 0.810 for $C_{h,udsc}$,
$C_{h,dk}$, and $C_{h,nl}$, and 0.870, 0.904 and 0.818 for
$C_{t,udsc}$, $C_{t,dk}$, and $C_{t,nl}$, respectively.

The uncorrected mean charged
multiplicity for all hadronic events was found
to be $\mh = 17.29 \pm 0.07$ tracks, while the mean charged
multiplicity
opposite tagged hemispheres was found to be $\mt = 9.28 \pm 0.09$
tracks. Combining these values with the $C_{i,j}$ via the
above relations
yields $\db = 1.94 \pm 0.30$(stat.) tracks.

We have investigated a number of systematic effects which may
bias the measured value of $\db$.
Dividing $\mh$ by the overall reconstruction
constant $C_{h,udscb} = 0.855$ provides a measurement of the
total hadronic multiplicity $\nhad = 20.21 \pm 0.08$(stat.).
This value is lower than the world average $20.95 \pm 0.20$ [9],
indicating that the detector
simulation overestimates the mean SLD tracking
efficiency by $\sim3.5\%$. We account for this by reducing all
reconstruction constants $C_{i,j}$ by this amount, leading to
a correction of $+ 0.10 \pm 0.10$ tracks in $\db$. We have
conservatively set the systematic error in the correction
to be equal to the size of the correction itself.

After correcting for overall tracking efficiency,
a comparison of the $p_{\perp}$ distribution between data and MC
shows good agreement for the untagged sample, but
an excess of
$\sim 15\%$ for data tracks opposite tagged hemispheres with
$p_{\perp}$ between $0.12$ and $0.50$ GeV/c,
accounting for $\sim 3 \%$ of all
reconstructed tracks in this sample. Since there are currently no
empirical constraints on the $p_{\perp}$ distribution
of non-leading tracks in $\zbbar$ events, we have assumed
that this excess is due to improper modelling of the non-leading
tracks by the JETSET MC, which to this point has been
tuned only to the global features of inclusive
$\zhad$ data.
We compensate
for this discrepancy by applying a further correction to
$\db$ of $+ 0.20 \pm 0.20$ tracks, where again we conservatively
assign an uncertainty equal in magnitude to the correction.
In addition, we have studied the behavior of $\db$ when numerous
other experimental parameters, such as tracking and event
selection requirements, were
varied over wide ranges. As a result of these studies, we assign
an additional systematic uncertainty of $\pm 0.15$ tracks due to
the uncertainty in charged-particle spectra modelling.

We have compared the fraction of tagged hemispheres
$f_t^{data} = 1829/10898 = 0.168 \pm 0.004$ to the MC expectation
$f_t^{MC} = 0.157$, assuming the world average value of
$R_b = 0.220 \pm 0.003$ [15].
If we conservatively assume that this difference
is due
entirely to extra
$Z^0 \rightarrow udsc$ contamination in the tagged sample,
the corresponding change in $\db$
is $0.21$ tracks. Since impact parameter reconstruction errors
tend to produce correlated changes in the $Z \rightarrow udsc$
and $\zbbar$ tagging efficiencies, the true uncertainty is somewhat
less than this. From MC studies of tracking errors which
produce the observed difference in $f_t$, we estimate the
systematic error due to the tagged sample purity to be
$\pm 0.15$ tracks.

An additional systematic error of $\pm 0.12$ tracks arises from
limited MC statistics.
Combining these uncertainties in quadrature, and
including the two corrections discussed above, we find
$$\db = 2.24 \pm 0.30(\rm{stat.})
\pm 0.33(\rm{syst.}) \; \rm{tracks}.$$
The effects of initial state radiation, and the $\sim 0.2$ GeV
difference between the mean cms energy of 91.55 GeV and the $Z^0$ peak,
are small, and no correction has been made.
Adding back in the world-average total hadronic multiplicity at the
$Z^0$ peak
$\nhad = 20.95 \pm 0.20$ [9] then yields
$$\nb = 23.19 \pm 0.30(\rm{stat.})
\pm 0.37(\rm{syst.}) \; \rm{tracks}.$$

To test the energy independence of the difference between
the total multiplicity in light quark events and the non-leading
multiplicity in $\zbbar$ events, we make use of lower cms energy
measurements
of the $e^+e^- \rightarrow b \overline{b}$ multiplicity from the PEP and
PETRA storage rings. Assuming the energy independence of the
decay multiplicity of $B$ hadrons produced in $e^+e^-$ annihilation,
it is equivalent to test the quantity $\Delta \nb \equiv \nb - \nuds$.
Results
for this quantity for the various lower cms energy experiments
are summarized in Ref.~[1].
Applying the
procedure presented in Ref.~[1] to the SLD measurement
 to remove the contribution from $\zccbar$, we
arrive at the result
$$\Delta \nb = 3.31 \pm 0.41(\rm{stat.})
\pm 0.53(\rm{syst.}) \pm 0.58(\overline{n}_c)
\; \rm{tracks},$$
where
we have constrained $\nc$ to lie
between $\nuds$ and $\nb$, yielding $\nc = 21.9 \pm 2.0$ tracks.

Figure 1 shows $\nhad$ and $\Delta \nb$ as
functions of cms energy. The $\Delta \nb$ data, with the additional
lever arm provided by the
SLD measurement, are seen to be
consistent with the hypothesis of energy independence, in marked
contrast to the steeply rising total multiplicity data [16].
Due to differing measurement techniques, results for
$\Delta \nb$ at PEP/PETRA energies are largely uncorrelated with
those at the $Z^0$ peak.
A linear fit to the $\Delta \nb$ data yields a slope
of $-1.0 \pm 1.1$ tracks/$\ln$(GeV), consistent
with $0$ at $0.9$ standard deviations.
Also
shown is
the perturbative QCD expectation for the value of
$\Delta \nb$.
Averaging the SLD result with previous measurements [1], we find that
$\Delta \nb^{comb} = 3.83 \pm 0.63$,
within 1.1 standard deviations of
the perturbative QCD
expectation of $5.5 \pm 0.8 \pm 1.2$(theory) [1].

The hypothesis of flavor-independent fragmentation [2,3], which
provides that $\nb(W) - \ndk(W) =
\nuds([1- \langle x_Q \rangle]W)$,
implies that
$\Delta \nb$ decreases with cms energy
in proportion to $\nuds(W)$ [1],
in contradiction with the perturbative QCD expectation.
Figure 2 shows a comparison between non-leading multiplicity
$\nb(W) - \ndk(W)$ and $\nuds([1- \langle x_Q \rangle]W)$, as a
function of non-leading energy
$[1- \langle x_Q \rangle]W$. When the SLD result is included,
a linear fit to the residuals (Fig. 2b) yields a slope of
$s = 1.91 \pm 0.65$ tracks/$\ln$(GeV), inconsistent with the
hypothesis of identical energy dependence ($s = 0.0$) at the level
of 2.9 standard deviations.

In conclusion, we have measured the difference in the mean charged
multiplicity between $\zbbar$ and $\zhad$ to be
$\db = 2.24 \pm 0.30(\rm{stat.}) \pm 0.33(\rm{syst.})$ tracks per event,
from which we calculate the multiplicity difference between $\zbbar$
and $\zuds$ to be
$\Delta \nb = 3.31 \pm 0.41(\rm{stat.})
\pm 0.53(\rm{syst.}) \pm 0.58(\overline{n}_c)$ tracks.
Comparing our measurement with similar results from lower energy
$e^+e^-$ annihilation data, we find no evidence that
$\Delta \nb$ depends on cms energy. This energy independence is
in agreement with the precise perturbative QCD expectation,
and indicates that QCD remains asymptotically free down to the
scale $M_b^2$. Our measured value
is in reasonable agreement with the less precise QCD prediction
that $\Delta \nb = \ndk - \nuds(\sqrt{e}M_Q)$. Including our
measurement, the cms energy dependence of the non-leading
multiplicity in $e^+e^-$ annihilation to $b$ quarks is
inconsistent with that of the hypothesis of flavor-independent
fragmentation at the level of 2.9 standard deviations.

We thank the personnel of the SLAC accelerator department
and the technical staffs of our collaborating institutions for their
outstanding efforts on our behalf.
We also thank Valery Khoze
for helpful and motivating discussions.
\vfill
\eject

\vskip .5\baselineskip

\noindent

\vskip .75\baselineskip

\vfill
\eject

\centerline{\bf References}

\vskip .5truecm

\point B. A. Schumm, Yu. L. Dokshitzer, V. A. Khoze, and
D. S. Koetke, Phys. Rev. Lett. {\bf 69}, 3025 (1992).

\point Mark II:
P. C. Rowson \etal, Phys. Rev. Lett. {\bf 54}, 2580 (1985).

\point A. V. Kisselev \etal, Z. Phys. C {\bf 41}, 521 (1988).

\point Mark II:
B. A. Schumm \etal, Phys. Rev. D {\bf 46}, 453 (1992).

\point SLD Design Report, SLAC--Report--273 (1984).

\point G. Agnew \etal, SLAC--PUB--5906 (1992).

\point E. Farhi, Phys. Rev. Lett {\bf 39}, 1587 (1977).

\point Impact parameter tagging with the SLD is discussed in detail
in K. Abe \etal, SLAC--PUB--6292, (August 1993).

\point Mark II: G. S. Abrams \etal, Phys. Rev. Lett. {\bf 64}, 1334
(1990); OPAL: P. D. Acton \etal, Z. Phys. C {\bf 53}, 539 (1992);
DELPHI: P. Abreu \etal, Z Phys. C {\bf 50} 185 (1991);
L3: B. Adeva \etal, Phys. Lett. B {\bf 259}, 199 (1991);
ALEPH: D. Decamp \etal, Phys. Lett. B {\bf 273}, 181 (1991).

\point W. Hollik, Fortschr. Phys. {\bf 38}, 165 (1990).

\point CLEO: R. Giles \etal, Phys. Rev. D {\bf 30},
2279 (1984); ARGUS: H. Albrecht \etal, Z. Phys. C
{\bf 54}, 13 (1992). Averaging the $\u4s$
multiplicity measurements from these sources yields
$\ndk = 10.88 \pm 0.20$.

\point ARGUS: H. Albrecht \etal, Z. Phys. C {\bf 58},
191 (1993).

\point T. Sj\"ostrand, Comput. Phys. Commun. {\bf43}, 367 (1987).

\point P. N. Burrows, Z. Phys. C {\bf 41}, 375 (1988),
OPAL: M. Z. Akrawy \etal, Z. Phys. C {\bf 47}, 505 (1990).

\point ALEPH, DELPHI, L3, OPAL: CERN--PPE--93--157
  (August 1993).

\point For a compilation of total and
non-leading multiplicity measurements in $e^+e^-$ annihilation,
as well as
heavy quark fragmentation parameters $\langle x_Q \rangle$,
see Ref.~[4].

\vfill
\eject


\centerline{\bf Figure Captions}

\vskip .5truecm

\noindent
Figure 1. Energy dependence of the total multiplicity [16]
and the
multiplicity difference
$\Delta \overline{n}_b$ [1,16]
between $e^+e^- \rightarrow b \overline{b}$ and
$e^+e^- \rightarrow uds$ events. A linear fit to the
energy dependence of $\Delta \overline{n}_b$
yields a slope of $s = -1.0 \pm 1.1$, consistent with the
hypothesis of energy independence ($s = 0.0$).
The horizontal lines are
the expected
value and $1\sigma$ range for
$\Delta \nb = \ndk - \nuds(\sqrt{e}M_b)$,
given by lower-energy total
multiplicity data in accordance with perturbative QCD (see text).

\noindent
Figure 2. a) Non-leading multiplicity $\nnl = \nb - \ndk$
in
$e^+e^- \rightarrow b \overline{b}$
{\it vs.}
non-leading energy $(1-\langle x_b \rangle)W$ [16].
The solid line is a fit [4] to $e^+e^- \rightarrow uds$
multiplicity as a function of W. The error on this fit (dotted lines)
is dominated by the uncertainty on the removal of the heavy
quark ($Q=c,b$) contribution to the measured $\nhad(W)$.
b) Residuals of a).

\vfill\eject

\bye